\documentclass{emulateapj}

\newcommand{\tab}[1]{Table~\ref{#1}}
\newcommand{\fig}[1]{Figure~\ref{#1}}
\newcommand{\eqn}[1]{equation~(\ref{#1})}

\newcommand{\powchi}{\ensuremath{7.8}}
\newcommand{\conchi}{\ensuremath{0.28}}
\newcommand{\aaaa}{\ensuremath{0.44\pm0.04}}
\newcommand{\bbbb}{\ensuremath{7.0\pm0.6}}
\newcommand{\cccc}{\ensuremath{3.1\pm0.4}}
\newcommand{\zfrac}{\ensuremath{1.3\pm0.4}}
\newcommand{\zdens}{\ensuremath{1.2}}
\newcommand{\tmerg}{\ensuremath{8.5}}
\newcommand{\delt}{\ensuremath{2.6}}

\begin{document}

\shorttitle{High Redshift Galaxy Mergers}
\shortauthors{Ryan Jr., et al.}

\title{Galaxy Mergers  at $z\!\gtrsim\!1$ in the HUDF:  Evidence for a
Peak in  the Major Merger  Rate of Massive  Galaxies\footnote{Based on
observations made  with the NASA/ESA Hubble  Space Telescope, obtained
from the Data Archive at  the Space Telescope Science Institute, which
is  operated  by  the  Association  of Universities  for  Research  in
Astronomy, Inc., under NASA contract NAS 5-26555.}}

\author{R. E. Ryan Jr.,}
\affil{Department of Physics, Arizona State University, Tempe, AZ 85281}
\email{russell.ryanjr@asu.edu}
\author{S. H. Cohen,  R. A. Windhorst,}
\affil{School of Earth and Space Exploration, Arizona State University, Tempe, AZ 85281}
\and
\author{J. Silk}
\affil{Department of Astrophysics, University of Oxford, Keble Road, Oxford OX1 3RH, UK}

\slugcomment{Accepted to ApJ}

\begin{abstract}
We  present a  measurement of  the galaxy  merger fraction  and number
density  from  observations  in   the  Hubble  Ultra  Deep  Field  for
$0.5\!\leq\!z\!\leq\!2.5$.  We  fit the combination  of broadband data
and  slitless spectroscopy  of 1308  galaxies with  stellar population
synthesis models  to select merging  systems based on stellar  mass of
$\geq\!10^{10}$~M$_{\odot}$.  When correcting for mass incompleteness,
the major merger fraction is not simply proportional to $(1+z)^m$, but
appears to  peak at $z_{\rm frac}\!\simeq\!\zfrac$.   From this merger
fraction,  we  infer  that   $\sim\!42$\%  of  massive  galaxies  have
undergone a major merger since $z\!\simeq\!1$.  We show that the major
merger number  density peaks at  $z_{\rm dens}\!\simeq\!\zdens$, which
marks  the epoch  where  major  merging of  massive  galaxies is  most
prevalent.  This  critical redshift is  comparable to the peak  of the
cosmic  star formation  rate density,  and occurs  roughly $\delt$~Gyr
earlier in  cosmic time  than the  peak in the  number density  of the
$X$-ray selected  active galactic nuclei.   These observations support
an indirect  evolutionary link between merging,  starburst, and active
galaxies.

\end{abstract}

\keywords{galaxies: evolution --- galaxies: formation --- galaxies: interaction}

\section{Introduction}\label{intro}

The hierarchical  formation scenario  suggests that galaxy  merging is
the driving  mechanism in galaxy  formation \citep[eg.][]{wf91,kwg93}.
Since  galaxy  merging  has  been  linked  to  numerous  astrophysical
processes, including star formation  and the fueling of a supermassive
black hole  (SMBH), which triggers  an active galactic  nucleus (AGN),
the   merger  history  of   galaxies  is   critical  for   a  complete
understanding of galaxy evolution.

The redshift  evolution of  the galaxy merger  fraction has  been well
studied   for  $z\!\lesssim\!1$,  and   is  roughly   proportional  to
$(1+z)^m$,         where        typically        $1\!\leq\!m\!\leq\!4$
\citep{zk89,burk94,cpi94,ye95,neu97,p97,lefe00,p02,lin04,xu,bell06,kart07}.
\citet{t77} predicted  the galaxy pair fraction to  be proportional to
the matter space density at low redshift, and \citet{c91} refined this
idea to show that the  power-law index should be somewhat sensitive to
the   Cosmological   Constant.    Based   on   $N$-body   simulations,
\citet{ber06}  argue that  while the  galaxy merger  rate per  halo is
increasing with redshift, the number of halos massive enough to host a
galaxy pair is  decreasing.  This balance will result  in a flattening
of the  high redshift pair  counts consistent with some  high redshift
observations  \citep[eg.][]{lin04}.   \citet{fak}  suggest  that  this
power-law form  is a particular  representation of a  universal merger
rate which  scales as a power-law  with the progenitor  mass ratio and
depends  only weakly  on the  halo mass.   Additionally, \citep{con06}
have  suggested that  the merger  fraction may  peak at  some critical
redshift.  However,  until now there have been  very few observational
studies for $z\!\gtrsim\!1$ to constrain the redshift of this possible
peak, and  hence the  detailed physics of  the galaxy merging  at high
redshift is still uncertain.

The  standard  hierarchical  formation  model  has  been  modified  to
incorporate    the    feedback    effects    of   a    central    SMBH
\citep[eg.][]{sr98}.   Hydrodynamical simulations have  suggested that
as  galaxies  merge,  they  evolve  into a  starburst  system,  and  a
substantial  amount   of  gas   is  eventually  funneled   onto  their
SMBHs. This triggers a visible  AGN typically 1--2~Gyr after the onset
of the merger \citep{sdh05}.  While previous authors have discussed an
evolutionary   link    between   mergers   and    starburst   galaxies
\citep[eg.][]{lt78,sm96,bell05},  this   ``duty  cycle''  of  merging,
followed by  a starburst, followed by  an AGN, has  only recently been
suggested \citep{hop06}.  This formation scheme implies that we should
observe similarities  in the  redshift evolution of  merging galaxies,
their star  formation history \citep[eg.][]{madau},  the mass assembly
in     galaxies     \citep[eg.][]{d03},     their     AGN     activity
\citep[eg.][]{ueda,hasi,sdss},   and  resulting   galaxy  morphologies
\citep[eg.][]{d95,abr96}.  Since these  other phenomena have been well
studied  at $z\!\gtrsim\!1$, we  desire corresponding  measurements of
galaxy merging  at comparable  redshifts, which is  the focus  of this
work.

While there may be some convergence on the theoretical front, there is
little  consensus  regarding  the  observational  connections  between
merging and  active galaxies.   In general, AGN  host galaxies  do not
exhibit strong  merger-type morphologies \citep[eg.][]{bah,shc,guy} or
show enhanced image asymmetry \citep{grog}.  Conversely, \citet{canal}
find evidence  for recent merger  activity in the quasar  host galaxy,
MC2~1635+119,  which  was  previously  classified  as  an  undisturbed
elliptical galaxy.  Some  of this disagreement may be  resolved in the
context  of  the  above  ``duty  cycle,''  wherein  the  morphological
indications of the recent merger have  subsided by the time the AGN is
observationally identified. This debate will surely continue owing, in
part,   to  the   difficulties   interpreting  morphological   studies
\citep[eg.][]{dep07}.

This work  is organized as  follows: In \S~\ref{data} we  describe the
dataset and  the mass estimates, \S~\ref{pairs} we  outline our merger
selection,  \S~\ref{evol} we  measure the  merger fraction  and number
density, \S~\ref{comp}  we compare this  study to published  work, and
conclude with  a general  discussion of the  result in the  context of
galaxy evolution in \S~\ref{discuss}.  We adopt the three-year results
from  the  {\it Wilkinson  Microwave  Anisotropy Probe}  \citep{wmap},
where     $\Omega_0\!=\!0.24$,     $\Omega_{\Lambda}\!=\!0.76$,    and
$H_0\!=\!100h$~km~s$^{-1}$~Mpc$^{-1}$    with    $h\!=\!0.73$.     All
magnitudes quoted herein are in the AB system \citep{abmag}.

\section{Data} \label{data}
The deep  fields observed  by the {\it  Hubble Space  Telescope} (HST)
provide    an   unprecedented   view    of   the    distant   universe
($z\!\gtrsim\!1$), and  are therefore the ideal  datasets for studying
the high redshift evolution of  the galaxy merger rate.  The GRism ACS
Program for Extragalactic  Science \citep[GRAPES;][]{nor} has provided
low-resolution  ($R\!\simeq\!100$) spectroscopy  of  1400~objects with
$z'\!\leq\!27.2$~mag     in     the     Hubble    Ultra Deep     Field
\citep[HUDF;][]{beck}.   From   these  data,  \citet{ryan07}  computed
spectrophotometric  redshifts for  1308 galaxies  at $z\!\lesssim\!5$.
When the  Balmer or  4000~\AA\ break occurs  in these  optical spectra
(for  $0.5\!\leq\!z\!\leq\!1.5$), the  standard deviation  in $(z_{\rm
phot}-z_{\rm spec})/(1+z_{\rm  spec})$ is 0.037, and it  is only 0.050
for  $z\!\geq\!1.5$,   which  make  these  among   the  most  accurate
photometric redshifts  computed using the  standard maximum-likelihood
technique \citep{hyperz} to date.

Since the  primary aim of  this work is  to measure the  high redshift
evolution of the  galaxy merger rate, we require a  set of criteria to
select  our  merger candidates.   While  this  has traditionally  been
performed using the rest-frame  luminosity, we will select our systems
based  on their mass.   Ideally, we  desire their  total gravitational
mass,    since   it    is   expected    to   drive    galaxy   merging
\citep[eg.][]{kb01}.    However,  since   this  mass   is  exceedingly
difficult to determine  for a large sample of  faint galaxies, we will
adopt the stellar mass as our selection quantity.

\begin{figure}
\epsscale{1.0}
\plotone{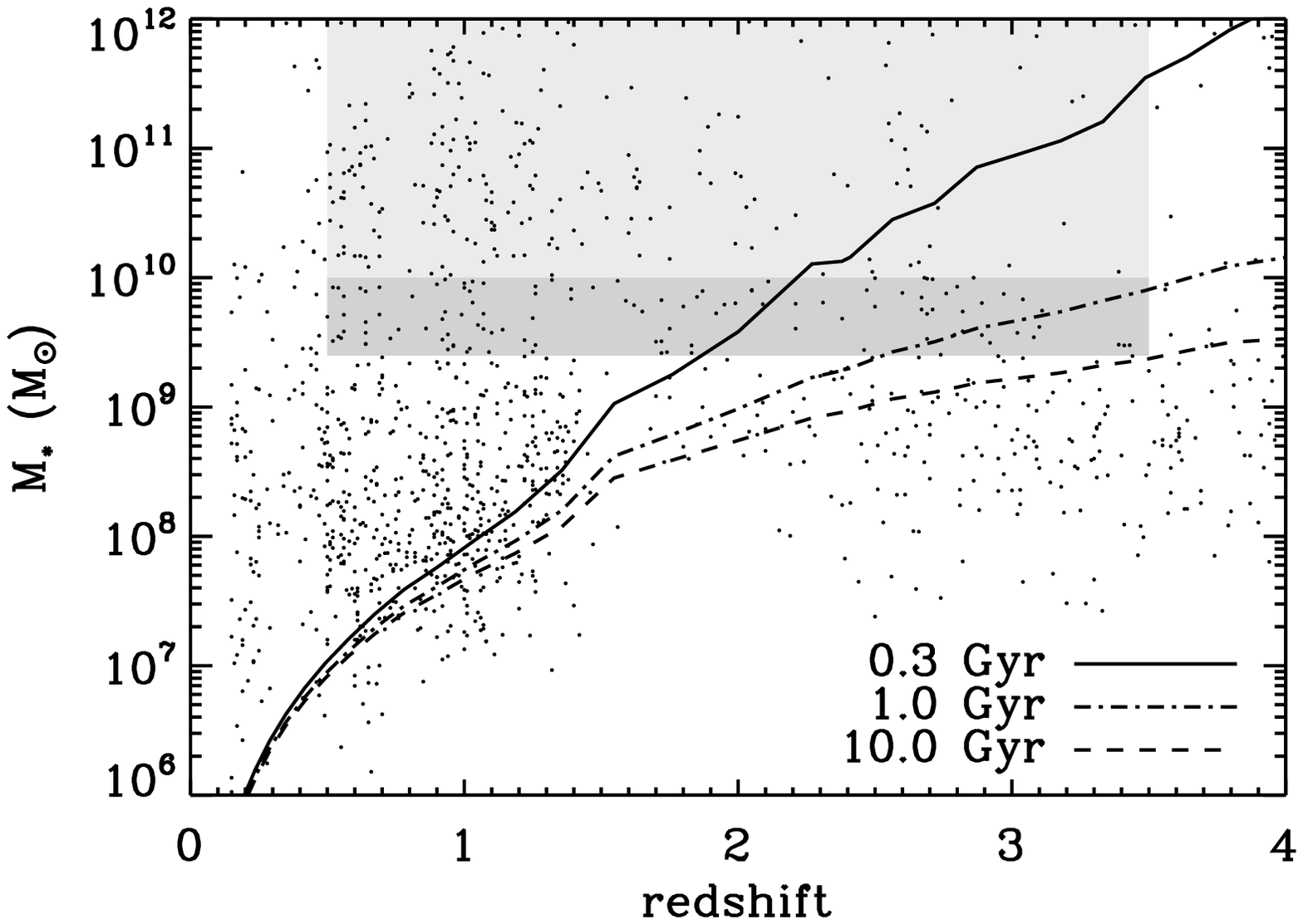}
\caption{The merger candidate selection.  We show the computed stellar
mass as  a function of  the spectrophotometric redshift.   The lightly
shaded region indicates the  volume-limited, mass selection region for
the  primary galaxies,  while  the darker  region  shows the  possible
selection  of  companion  galaxies.   In total,  $\sim\!95$\%  of  the
galaxies   with  $z'\!\leq\!27.2$~mag  identified   in  the   HUDF  by
\citet{beck} are represented here (see \fig{numcts}) with 230 of these
meeting  the  selection  criteria  discussed in  \S~\ref{pairs}.   The
various  lines show the  mass of  a maximally  old, passively-evolving
stellar population with an $e$-folding time in the star formation rate
given in the lower-right corner.}\label{massfig}
\end{figure}

To determine  the required  stellar masses of  our sample, we  fit the
combination of GRAPES spectra  and broadband observations in the $B$-,
$J$-, and $H$-bands with a grid of stellar population synthesis models
\citep[][BC03   hereafter]{bc03}.   The   masses  are   determined  by
minimizing  the  reduced $\chi^2$  between  the spectrophotometry  and
these models, following the  method of \citet{pap}.  By parameterizing
the    star     formation    history    of     these    galaxies    as
$\Psi(t)\!=\!\Psi_0e^{-t/\tau}$,  our  model  grid consists  of  three
parameters: the extinction in  the restframe $V$-band, the $e$-folding
time in the star formation history, and stellar population age.  If we
assume  a  Salpeter  initial  mass  function  \citep{salp}  and  solar
metallicity,     then    the     parameter    space     consists    of
$\simeq\!4\times10^4$~independent  models.  By  varying the  values of
the grid points, we estimate  the uncertainty on the stellar masses to
be $\lesssim\!15$\%.  In \fig{massfig}, we  show the stellar mass as a
function of  redshift for these  galaxies.  The different  tracks show
the mass of a maximally  old stellar population for the various values
of $\tau$ listed  in the lower-right corner.  The  shaded region shows
the  mass selection for  the primary  galaxies, which  is $\sim\!95$\%
complete  for  $z'\!<\!27.2$~mag.    The  darker  area  indicates  the
selection region for the lower mass companion galaxies to be discussed
below.

\section{Identification of Merging Galaxies}\label{pairs}

Since we have reasonable redshift information for the entire sample of
galaxies,  it is  natural that  we  adopt the  dynamically close  pair
method \citep[eg.][]{cpi94}.  However,  since our redshift uncertainty
is  roughly  a  factor of  $10$  larger  than  that of  a  traditional
spectroscopic survey, our  approach will be somewhat akin  to the more
common  pair count analysis  \citep[eg.][]{zk89}.  This  technique has
several key virtues not present in other approaches: it has simple and
well understood statistics to  compute \citep{abr99} and does not rely
on calibration datasets \citep{dep07}.  \citet[][P2000 hereafter]{p00}
demonstrate that the  galaxy merger rate (${\cal R}_{\rm  mg}$) can be
computed from the galaxy merger  or pair fraction ($f$) and the galaxy
number density ($n$)  as ${\cal R}_{\rm mg}\!=\!fn\tau_{\rm mg}^{-1}$,
where $\tau_{\rm mg}$ is the typical merger timescale.

\subsection{Pair Criteria} \label{crit}
Using the redshift catalog of \citet{ryan07}, we require the following
four criteria  to be satisfied to  identify a galaxy pair  as a merger
system: (1) at least one neighboring galaxy $\leq$20~$h^{-1}$~kpc away
to ensure that close galaxy pairs will merge \citep[eg.][]{mgc,bell06}
on  a timescale  $\lesssim\!0.5$~Gyr  \citep[eg.][]{kb01,p02}; (2)  we
require the mass ratio of  primary galaxy to the putative companion to
be $\leq\!4\!:\!1$, which is consistent  with a flux ratio between the
galaxies of  $\lesssim\!1.5$~mag; (3) we  require the stellar  mass of
the primary galaxy to  be $M_*\!\geq\!10^{10}$~M$_{\odot}$; and (4) we
require the redshifts of the primary and the companion to agree within
the 95\%  confidence interval  of the spectrophotometric  redshifts of
\citet{ryan07}.

\subsection{Accounting for the Flux and Mass Limit} \label{fluxlim}

\begin{figure}
\epsscale{1.0}
\plotone{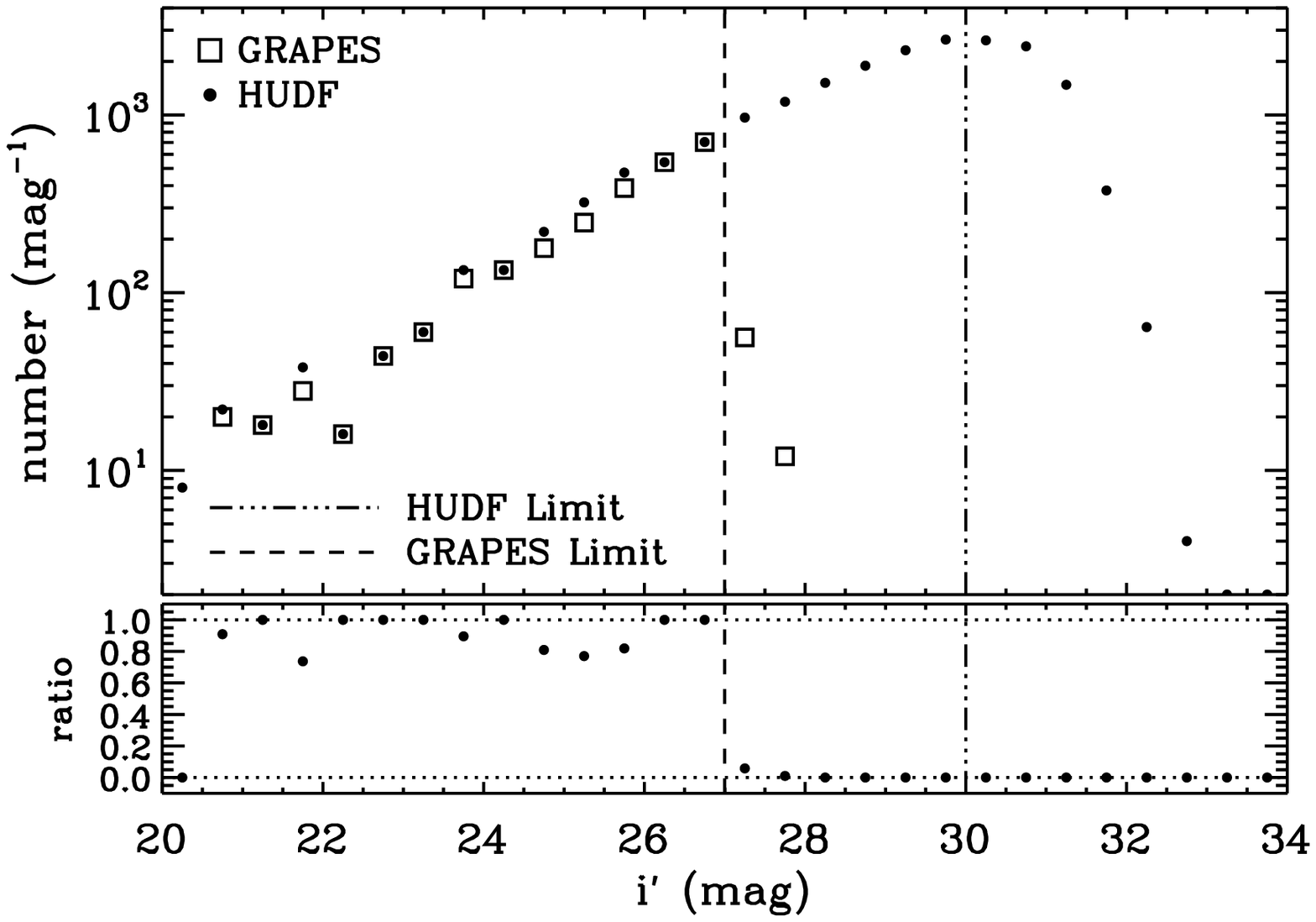}
\caption{The $i'$-band  number counts.  Since the  GRAPES objects were
selected  from  the  HUDF  to  have $z'\!\leq\!27.2$~mag,  it  is  not
surprising that  $\geq\!95$\% complete at  these flux levels.   In the
lower panel,  we show  the ratio  of the number  counts of  the GRAPES
survey to all objects in the HUDF.}
\label{numcts}
\end{figure}

\begin{figure}
\epsscale{1.0}
\plotone{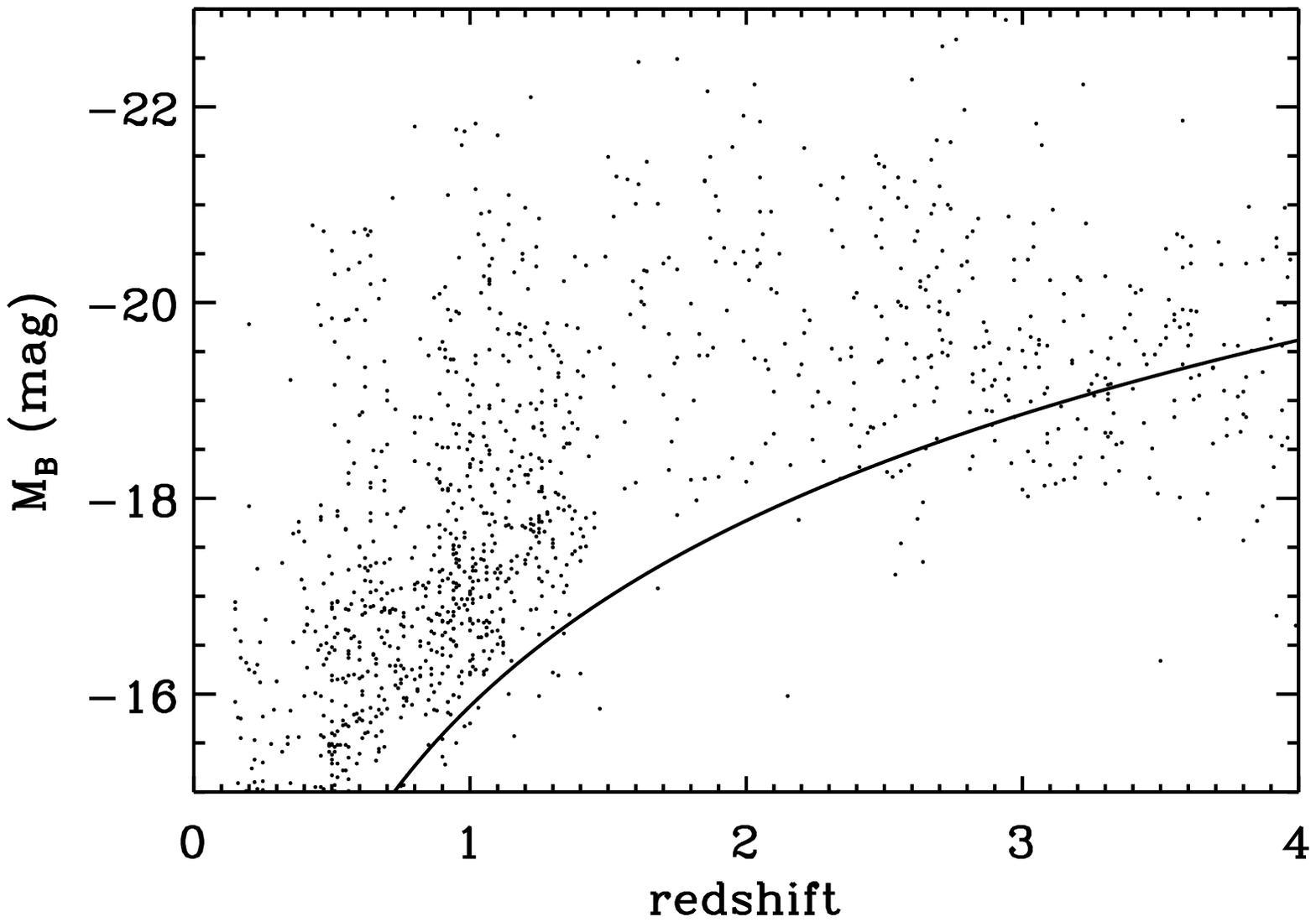}
\caption{The $B$-band absolute magnitude  shown for comparison and not
used in  the merger  selection.  The solid  line shows the  contour of
constant   apparent  magnitude  of   $z'\!\simeq\!27.2$~mag  following
\citet{driv}.     The   majority   of    our   galaxies    have   ages
$\lesssim\!500$~Myr, which implies  their median stellar mass-to-light
ratio  is  $\Upsilon_{*,B}\!\simeq\!0.4$~M$_{\odot}$~L$_{\odot}^{-1}$.
Therefore, the mass  limit of $M_*\!\geq\!10^{10}$~M$_{\odot}$ roughly
corresponds  to  $M_B\!\leq\!-20.5$~mag, which  is  comparable to  the
luminosity limit imposed  by various authors \citep[eg.][]{p97,lefe00}
at lower redshifts.  }\label{amag}
\end{figure}

While the  GRAPES survey  is complete to  very faint flux  levels (see
\fig{numcts}), we may expect it  to be volume-limited at high redshift
and  that  no corrections  to  the  pair  statistics are  needed.   To
investigate this notion, we show  the absolute $B$-band magnitude as a
function of  the spectrophotometric  redshift in \fig{amag}  where the
solid line shows  our flux limit of $z'\!\simeq\!27.2$~mag  for a flat
spectrum in $f_{\nu}$ following \citet{driv}.  While this figure shows
the  expected volume-limited behavior  \citep[eg.][]{mgc}, there  is a
more subtle question to address  since this is a mass-selected sample:
Is  there a population  of galaxies  which are  fainter than  our flux
limit  and more  massive than  our mass  limit? If  such  a population
exists, then our mass selection  would be incomplete in these galaxies
\citep{d03}.  At the  redshifts of interest, such objects  must have a
relatively        high        stellar       mass-to-light        ratio
($\Upsilon_{*,B}\!\gtrsim\!2$~M$_{\odot}$~L$_{\odot}^{-1}$)        and
therefore be very red in the HST images.

\begin{figure}
\epsscale{1.0}
\plotone{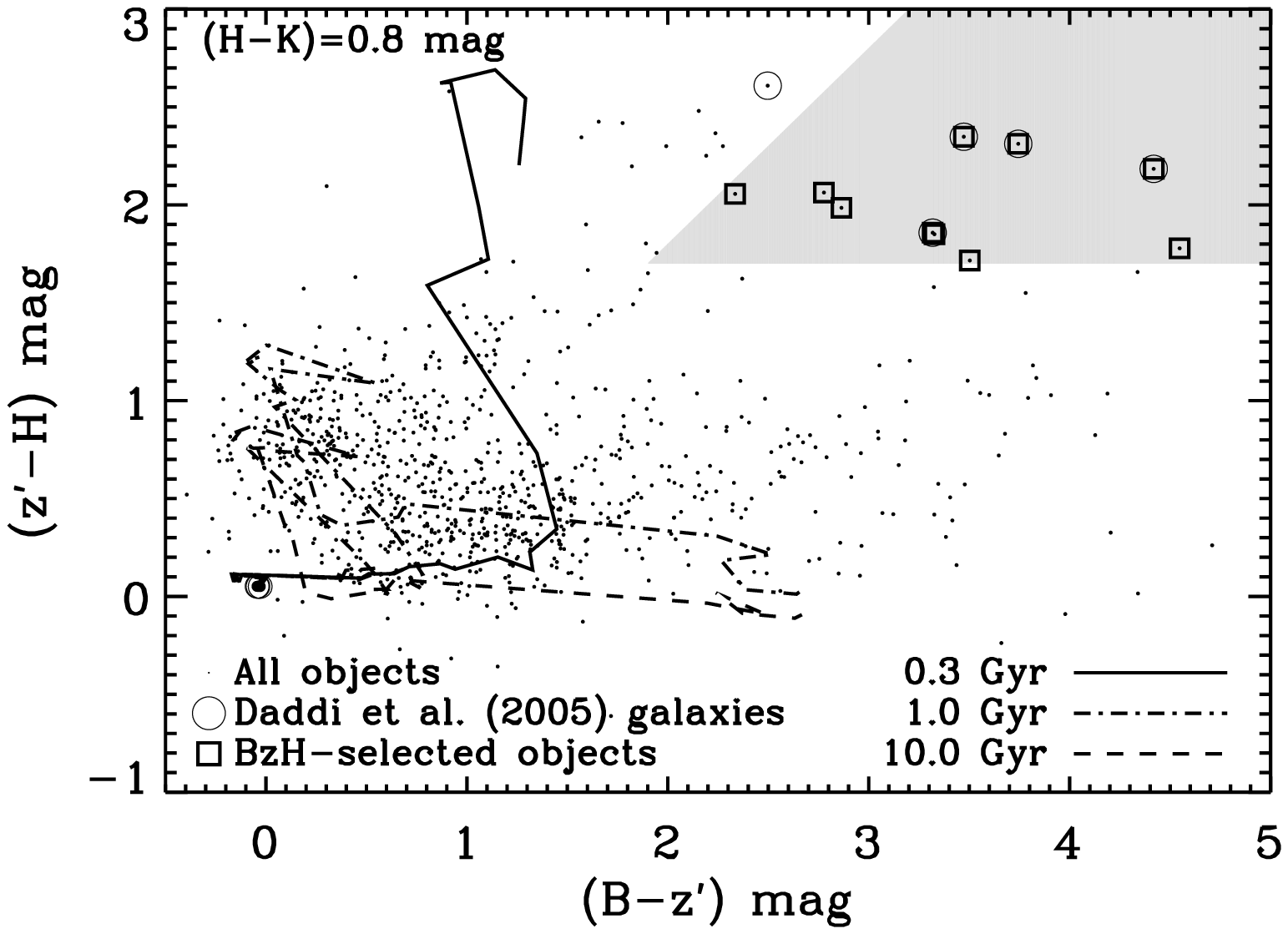}
\caption{The  $BzH$  selection   for  passively  evolving,  burst-type
galaxies.   By   assuming  $(H-K)\!\simeq\!0.8$~mag  \citep{fs04},  we
perform a similar analysis of  \citet{dad05} on the much deeper NICMOS
imaging in the HUDF to  estimate the range of star formation histories
applicable to these redshifts.  The shaded region represents the $BzH$
selection discussed in \S~\ref{fluxlim}  and, as in \fig{massfig}, the
different  lines represent  maximally old,  passively-evolving stellar
populations, denoted  by the $e$-folding time in  their star formation
rate  in the  lower-right.  The  ``bullseye'' marks  the birth  of the
populations at  $z_{\rm form}\!\rightarrow\!\infty$.  As  discussed in
\S~\ref{fluxlim},  maximally old  galaxies  with $\tau\!\leq\!0.3$~Gyr
are exceedingly  rare and would  appear in the shaded,  $BzH$ polygon.
Therefore, when correcting our pair counts following the P2000 method,
we    adopt     BC03    models    appropriate     for    our    sample
($\tau\!\geq\!0.3$~Gyr).}\label{colcol}
\end{figure}

To find these dark, red galaxies at $1.5\!\lesssim\!z\!\lesssim\!2.4$,
\citet{dad04}  have  developed  a  technique  based  on  the  observed
$(B-z')$ and $(z'-K)$ colors (the $BzK$ method).  When this is applied
to the HUDF, \citet{dad05} identify only seven old, passively-evolving
galaxies with $K\!<\!23$~mag.  They argue that these systems must have
been formed in a  quick burst ($\tau\!\leq\!0.3$~Gyr), since they emit
little  restframe, ultraviolet  flux.   While these  objects meet  the
above mass-to-light  criteria, they are  brighter than our  flux limit
and are included  in our redshift survey.  Therefore,  to estimate the
number  of galaxies  satisfying this  mass-to-light criterion  and are
fainter than  our flux  limit, we must  perform a similar  analysis on
deeper data. Since  there is no sufficiently deep  $K$-band imaging in
the HUDF, we  must use the NICMOS imaging  in the $H$-band \citep{nic}
and assume  that these galaxies  typically have $(H-K)\!\sim\!0.8$~mag
\citep{fs04}.  In \fig{colcol},  we show the corresponding color-color
space for  the $BzH$ selection for  the objects in  the NICMOS imaging
with  fluxes from  \citet{coe}.  The  shaded polygon  shows  the $BzH$
selection  region  defined  by $(z'-H)-(B-z')\!<\!-0.2-(H-K)$~mag  and
$(z'-H)\!>\!2.5-(H-K)$~mag.   The open  squares represent  all objects
satisfying these $BzH$  criteria while the open circles  show the five
objects of \citet{dad05} which are in this portion of the HUDF.  As in
\fig{massfig}, the various tracks  show BC03 stellar population models
as indicated  in the  lower-left.  Since we  recover 4/5  objects from
\citet{dad05}, we  conclude that this is  a viable method  to find the
maximally   old,  passively-evolving   galaxies   which  satisfy   the
mass-to-light requirement.   There are only 9~galaxies (or  1\% of all
objects  with  $z'\!\lesssim\!28.5$~mag)  which  are  expected  to  be
burst-type   systems  \citep[or   a  $\tau\!\leq\!0.3$~Gyr;][]{dad05}.
However, the  abundance of galaxies with masses  and colors comparable
to    the    maximally    old,    passively-evolving    models    with
$\tau\!\geq\!0.3$~Gyr in \fig{massfig}  and \fig{colcol} suggests that
corrections to the pair statistics are necessary.

To  correct the  observed pair  counts, we  will adopt  the  method of
P2000,  and  for brevity  only  highlight  our  modifications for  the
mass-selection and  give the essential details.  In  the P2000 method,
one  computes the  maximal $k$-correction  from  a set  of galaxy  SED
templates to  derive the low  luminosity limit when  selecting objects
based on absolute magnitudes  \citep[see Figures~1 and 2 from][]{p02}.
However, since  we select candidates  based on their stellar  mass, we
must   determine  the   corresponding  maximal   mass   limit.   While
\citet{p02} adopt  the empirical  templates of \citet{cww},  we choose
the stellar population synthesis models of BC03.  As in \S~\ref{data},
we      parameterize     the      star      formation     rate      as
$\Psi(t;\tau)\!\propto\!e^{-t/\tau}$            and           restrict
$\tau\!\geq\!0.3$~Gyr,  since objects  with  $\tau\!\leq\!0.3$~Gyr are
extremely rare (as  discussed above).  Since the aim  is to generate a
population of maximal stellar  mass, we set $t\!=\!T(z)$, where $T(z)$
is the age of the Universe at redshift $z$.

We define  the stellar mass limit  to be the more  massive between the
fiducial limit ($M_{*,0}$) and the maximal mass model ($M_*(z;\tau)$):
\begin{equation}
M^{\rm lim}_*(z)=\max{[M_{*,0},M_*(z;\tau)]}.
\end{equation}
With this mass selection, we compute the selection function to correct
the counts of the flux-limited survey as
\begin{equation}
S_{N}(z)=\frac{\int_{M_{*,0}}^{M^{\rm lim}_*(z)}\Phi(M_*,z)dM_*}{\int_{M_{*,0}}^{\infty}\Phi(M_*,z)dM_*},
\end{equation}
where $\Phi(M_*,z)$  is the galaxy  mass function (not to  be confused
with the luminosity function).  \citet{elsn} show that the galaxy mass
function  can be well  modeled as  the Schechter  function \citep{sch}
with  $M^*_*\!\sim\!10^{11.5}$~M$_{\odot}$  and  $\alpha\!\sim\!-1.36$
for $z\!\lesssim\!5$.   Finally, the optimal  weights to apply  to the
pair  statistics  will  be $w_N(z)\!\propto\!S_N(z)^{-1}$,  where  the
various remaining  terms (such  as boundary effects  and spectroscopic
completeness) are computed following \citet{p02}.  Using these optimal
weights, the number  of close companions for the  $i$th primary galaxy
is computed as $N_{c,i}\!=\!\sum_j  w_{N_2}(z_j)$, where $j$ runs over
the companions which satisfy  the above merger criteria.  Finally, the
mean number of close companions per primary galaxy is found as
\begin{equation}
N_c=\frac{\sum_i w_{N_1}(z_i)N_{c,i}}{\sum_i w_{N_1}(z_i)},
\end{equation}
following the notation of \citet{p02}.

\subsection{Error Budget}\label{budget}

While it  may be  appropriate to assume  Poisson uncertainties  on the
pair counts,  we require a  more detailed analysis to  incorporate the
effects of  the redshift uncertainties.  Despite  having redshifts for
each component of the merging pair, the relatively large uncertainties
on the  proper distance ($\delta d_p\!\lesssim\!500$~Mpc)  can lead to
false  positives.   To  estimate  the  magnitude of  this  effect,  we
randomize the  redshifts and masses in  such a manner  to preserve the
selection discussed  above and recompute  the pair counts.   In $10^3$
realizations of the redshift catalog, we estimate that $\lesssim\!1.5$
pairs  per redshift bin  may arise  from superpositions  identified by
erroneous  redshifts.   This  approach  allows  us  to  bootstrap  the
uncertainties on  the pair  counts to include  the effects  of Poisson
shot noise, chance superpositions, and redshift uncertainties.

Since   the  HUDF  covers   $\simeq\!3\farcm3\!\times\!3\farcm3$,  the
uncertainties resulting  from cosmic variance must  also be addressed.
By   assuming   the   correlation   function  of   the   galaxies   is
$\xi(r)\!\propto\!(r/r_0)^{-\gamma}$,   then   the   relative   cosmic
variance     in     the      number     counts     is     given     as
$\sigma_v^2\!=\!1.45(V/r_0^3)^{-\gamma/3}$   for   cell  volume   $V$,
correlation  length  $r_0$,  and  $\gamma\!=\!1.8$  \citep{som}.   For
comparable sampled volumes,  \citet{beck} estimate the cosmic variance
to be $\sigma_v\!\simeq\!0.17$ for the HUDF cell geometry.

\section{Redshift Evolution} \label{evol}
\begin{figure}
\epsscale{1.0}
\plotone{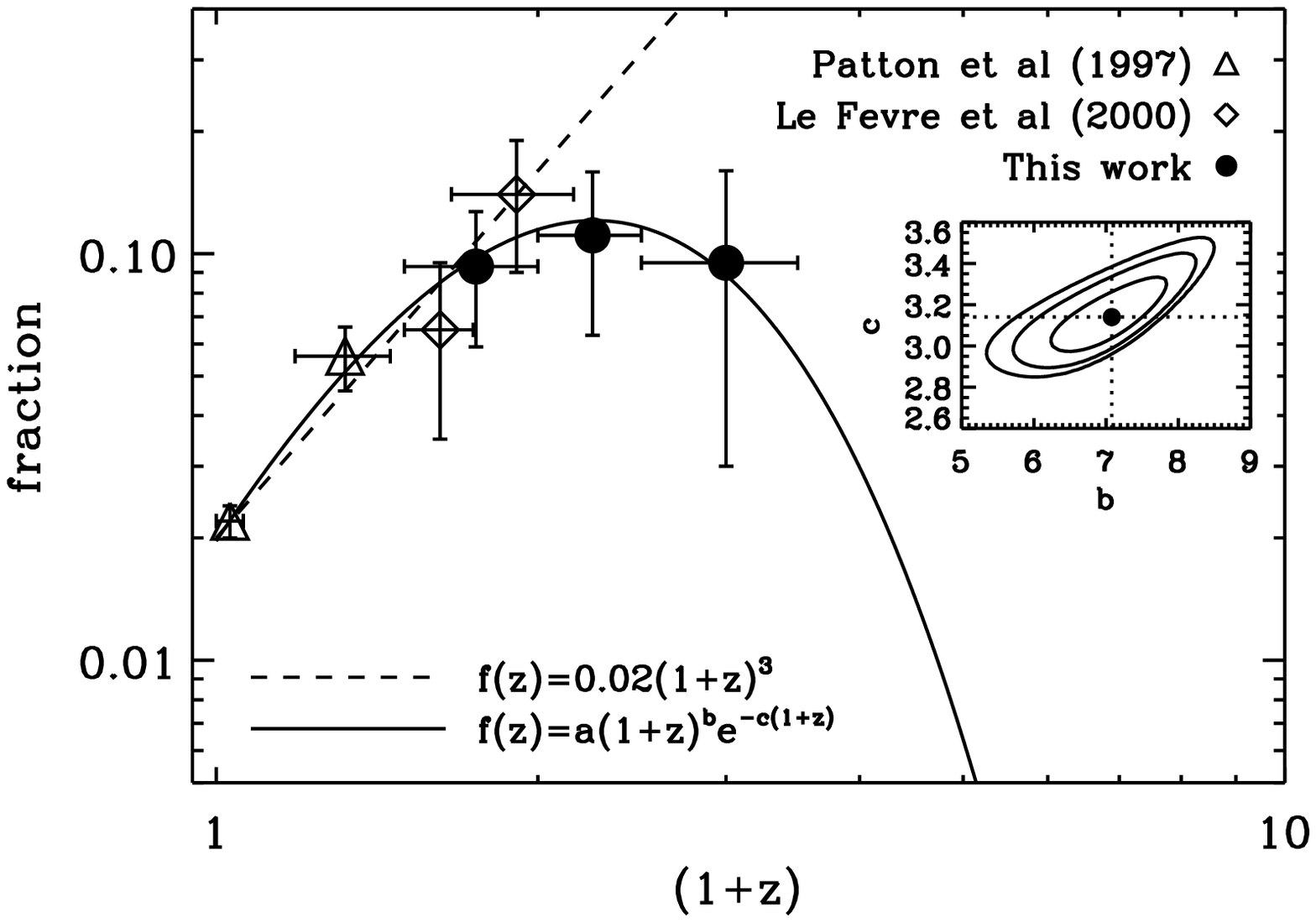}
\caption{The observed  galaxy merger  fraction. The results  from this
work  are  indicated by  filled  circles,  while  the observations  of
\citet{p97}  and  \citet{lefe00}  are  shown  by  open  triangles  and
diamonds,  respectively.  The  dashed line  shows the  common  form of
$f(z)\!=\!0.02\times(1+z)^3$,  which fails  to fit  our  high redshift
observations.   However,   the  model  $f(z)\!=\!a(1+z)^be^{-c(1+z)}$,
shown as the solid line, can reproduce all the observations reasonably
well ($\chi^2_{\nu}\!=\!\conchi$).   This peak in  the merger fraction
at $z_{\rm frac}\!\simeq\!\zfrac$ may arise from the strict mass limit
($M_*\!\geq\!10^{10}$~M$_{\odot}$)  and  mass ratio  ($\leq\!4\!:\!1$)
imposed on our sample and is  not due to flux incompleteness given the
unparalleled  depth  of  this   survey  (see  \fig{massfig})  and  the
corrections applied in \S~\ref{fluxlim}.}\label{fracfig}
\end{figure}

In \fig{fracfig},  we show the  resulting galaxy merger fraction  as a
function of $(1+z)$ as  filled circles with uncertainties that reflect
the  contributions   from  chance  superpositions,   cosmic  variance,
redshift uncertainty,  and Poisson  shot noise.  Since  these galaxies
are typically very young  (an age $\lesssim\!500$~Myr as inferred from
the  SED fitting  described  in \S~\ref{data}),  their median  stellar
mass-to-light     ratio    in     the     restframe    $B$-band     is
$\Upsilon_{*,B}\!\simeq\!0.4$~M$_{\odot}$~L$_{\odot}^{-1}$,         and
therefore  the   mass  limit  of  $M_{*,0}\!\geq\!10^{10}$~M$_{\odot}$
roughly   corresponds  to   $M_B\!\lesssim\!-20.5$~mag.    Since  this
luminosity  limit  is  similar  to  that imposed  by  \citet{p97}  and
\citet{lefe00},  we show  those  measurements for  comparison as  open
triangles   and   diamonds,   respectively.    We  show   a   standard
parameterization  of $f(z)\!=\!0.02\times(1+z)^3$  as  a dashed  line.
While  this   model  can  fit  the   low  redshift  ($z\!\lesssim\!1$)
observations,   it   fails    to   reproduce   our   higher   redshift
($z\!\gtrsim\!1$) data.  Therefore, we  also show the empirical formed
proposed by \citet{con06} of
\begin{equation}\label{fmod}
f(z)=a(1+z)^be^{-c(1+z)},
\end{equation}
as a  solid line.  Using  a standard least-squares technique,  we find
$a\!=\!\aaaa$,      $b\!=\!\bbbb$,      and     $c\!=\!\cccc$      for
$\chi^2_{\nu}\!=\!\conchi$.   This  model  has  a maximum  at  $z_{\rm
frac}\!=\!(b/c)-1\!=\!\zfrac$,  when  accounting  for  the  correlated
uncertainties on $b$  and $c$.  In the inset, we  show the 68\%, 95\%,
and  99\% confidence intervals  in the  $b$-$c$ plane  for the  fit to
these data.  The  peak in the fraction is likely  a consequence of the
strict mass  limit and mass  ratio imposed, and  not a result  of flux
incompleteness (see \fig{massfig}).  Our observations can rule out the
standard     power-law      model     which     has      a     reduced
$\chi_{\nu}^2\!=\!\powchi$.

\begin{figure}
\epsscale{1.0}
\plotone{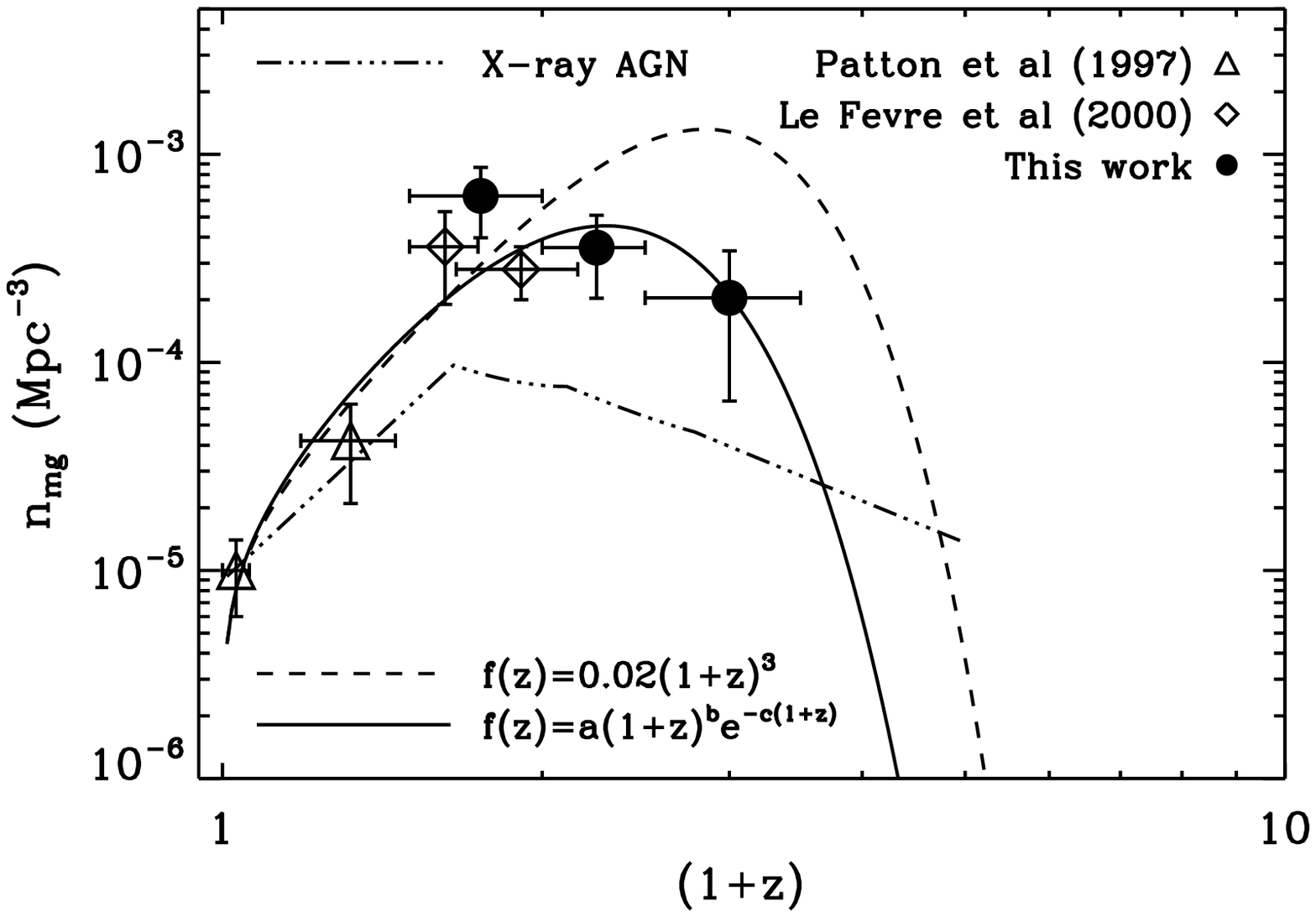}
\caption{The   observed   galaxy  merger   number   density.   As   in
\fig{fracfig},  the results  from this  work are  indicated  by filled
circles, and  the observations  of \citet{p97} and  \citet{lefe00} are
shown as  open triangles and diamonds, respectively.   The dotted line
shows  the total  density of  AGN  with $L_X\!>\!10^{42}$~erg~s$^{-1}$
selected  in the  soft  (0.5--2~keV) $X$-ray  band \citep{hasi}.   The
number    density    of    merging    galaxies   peak    at    $z_{\rm
dens}\!\simeq\!\zdens$ while  the AGN number density  peaks at $z_{\rm
AGN}\!\simeq\!0.65$.   This difference  in redshift  corresponds  to a
cosmic time difference of  $\simeq\!2.5$~Gyr.  A comparable time delay
between  the  merging  and  the   AGN  phase  has  been  suggested  by
simulations \citep{sdh05}.   Moreover, the  ratio of the  densities at
their peaks  would suggest that as  much as $\sim\!20$\%  of the major
merging  of  massive  galaxies  will  evolve  into  $X$-ray  AGN  with
$L_X\!>\!10^{42}$~erg~s$^{-1}$ \citep[eg.][]{dad07}.}\label{densfig}
\end{figure}

While the merger fraction  may provide excellent information on galaxy
evolution,  the galaxy  merger number  density is  more  relevant when
comparing  to  other  extragalactic  observations.   From  the  merger
fraction ($f$), we compute the merger number density as
\begin{equation} \label{denseqn}
n_{\rm mg}=\frac{f(z)N(z)}{\int_{z-\Delta z}^{z+\Delta z}dV},
\end{equation}
where $N(z)$ is the  completeness corrected redshift distribution from
the  \citet{ryan07} catalog. Then  the merger  rate is  ${\cal R}_{\rm
mg}\!=\!n_{\rm mg}\tau_{\rm  mg}^{-1}$ (P2000).  In  \fig{densfig}, we
show  the merger  number density  for the  same data  as  discussed in
\fig{fracfig}.   As   a  solid  line,  we  show   empirical  model  of
\eqn{fmod},  converted   to  a  number  density   using  the  redshift
distribution  of equation~(23)  in \citet{ben00}.   The  merger number
density peaks at a $z_{\rm dens}\!\simeq\!\zdens$, which is a slightly
lower  redshift than  the peak  in the  merger fraction.   Finally, we
tabulate  our measured galaxy  merger fraction  and number  density in
\tab{merg}.

\begin{table}
\caption{Galaxy Merger Results}\label{merg}
\begin{tabular*}{0.48\textwidth}
  {@{\extracolsep{\fill}}ccc}
\hline
\hline
\multicolumn{1}{c}{$z$} & \multicolumn{1}{c}{$f(z)$} & \multicolumn{1}{c}{$n_{\rm mg}$\tablenotemark{$\dagger$}}\\
\multicolumn{1}{c}{$ $} & \multicolumn{1}{c}{$ $} & \multicolumn{1}{c}{(10$^{-4}$~Mpc$^{-3}$)}\\
$ $&$ $&$ $\\
\hline
0.75$\pm$0.25 & 0.09$\pm$0.03 & 6.31$\pm$2.34\\
1.25$\pm$0.25 & 0.11$\pm$0.04 & 3.56$\pm$1.52\\
2.00$\pm$0.50 & 0.09$\pm$0.06 & 2.04$\pm$1.39\\
\hline
\tablenotetext{$\dagger$}{Computed from the merger fraction by \eqn{denseqn}.}
\end{tabular*}
\end{table}
\enddata

\section{Comparison to Previous Work}\label{comp}
For  low to  moderate  redshifts ($z\!\lesssim\!1$),  there have  been
numerous measurements of the galaxy pair fraction.  While many studies
disagree on  the precise values  of the pair fraction,  they generally
agree that  the fraction rises  from $\sim\!2$\% at  $z\!\simeq\!0$ to
$\sim\!15$\% at  $z\!\simeq\!1$ as a power-law in  $(1+z)$.  Where our
observations overlap with previous work,  we find a merger fraction of
$15\pm5$\%  at  $z\!\simeq\!1$,  which   is  in  good  agreement  with
published values \citep[eg.][]{lefe00,lin04,kart07}.

At higher  redshift ($z\!\gtrsim\!1$), there  are no other  studies of
merging using  the pair fraction,  due to the difficulty  in acquiring
sufficiently deep redshift datasets.   However, if the redshift of the
primary galaxy can be obtained, then a morphological approach can be a
viable method  for determining  the merger status.   \citet{con03} has
shown  a roughly  power-law  form to  $z\!\simeq\!3$  using the  $CAS$
(concentration,    asymmetry,    clumpiness)   morphological    system
\citep{ber00,con03a}.   When  these   measurements  are  coupled  with
restframe colors, the refined  merger fraction is substantially higher
than our  observations \citep{con06}.  However,  \citet{dep07} discuss
some of the limitations of this system and caution that merger studies
using image asymmetry require careful interpretation.

Based  on the  matched Two  Micron All  Sky  Survey (2MASS)/Two-Degree
Field  Galaxy   Redshift  Survey  \citep[2dFGRS;][]{cole},  \citet{xu}
identify  38~merging  systems  with  similar close  pair  criteria  at
$0.01\!\lesssim\!z\!\lesssim\!0.07$.   By adopting a  standard stellar
mass-to-light      ratio     in      the      restframe     $K_s$-band
($\Upsilon_{*,K_s}\!=\!1.32$~M$_{\odot}$~L$_{\odot}^{-1}$),       these
galaxies  are  selected  to have  $M_*\!\geq\!10^{10.25}$~M$_{\odot}$.
While they restricted their work to a smaller range of mass ratios for
their systems  ($\delta K_s\!\leq\!1$~mag), they find  a pair fraction
of $1.70\pm\!0.32$\%, consistent with  other values at these redshifts
\citep[eg.][]{p97}.

\section{Discussion}\label{discuss}

Using the  deepest imaging and  spectroscopic data available,  we show
that the fraction and number  density of major mergers between massive
galaxies peak at relatively low  redshifts. These peaks are likely not
an artifact of the data nor the measurement process, given the careful
attention  to the mass  limits, spectrophotometric  redshift accuracy,
accounting of measurement uncertainties, unprecedented flux limit, and
the  good agreement  with  published  data in  the  redshift range  of
overlap.

Since  the   high  redshift  portion  of  our   observed  merger  rate
dramatically differs  from what  is found by  numerous authors,  it is
interesting to explore some of  the consequences which arise from this
difference.   In  particular,  one  implication  of  our  merger  rate
concerns the fraction of galaxies which have undergone a major merger,
or the remnant fraction $f_{\rm  rem}$.  By assuming a merger fraction
of $f(z)\!=\!0.011\times(1+z)^{2.3}$ and a merger timescale $\tau_{\rm
mg}\!=\!0.5$~Gyr, \citet{p02} find that $f_{\rm rem}\!\simeq\!15$\% of
galaxies with $-21\!\leq\!M_B\!\leq\!-18$~mag have experienced a major
merger   since  $z\!\simeq\!1$.    This  is   in  contrast   with  the
semi-analytic models  of \citet{del},  who argue that  $\sim\!50$\% of
stars which  comprise elliptical galaxies assemble into  a single dark
halo between  $0\!\leq\!z\!\leq\!1$ by  major merging.  When  we apply
this   analysis\footnote{Our   remnant    fraction   is   found   from
equation~(18)  of \citet{p02}.}   to our  slightly larger  and steeper
merger rate, we find that $f_{\rm rem}\!\simeq\!42$\% of galaxies with
$M_*\!\geq\!10^{10}$~M$_{\odot}$ have been  involved in a major merger
since  $z\!=\!1$.   Moreover, since  our  merger  rate  peaks at  some
critical  redshift,  the  fraction   of  massive  galaxies  that  have
undergone   a   major   merger   tends   to  a   limiting   value   at
$z\!\rightarrow\!\infty$  of  $f_{\rm  rem}^{\rm  lim}\!\simeq\!62$\%,
whereas  a strict power-law  form increases  to 100\%.   This suggests
that $(1-f_{\rm rem}^{\rm lim})\!\simeq\!38$\% of massive galaxies may
have never undergone a major merger in their lifetime, but does not 
preclude a series of minor mergers.

The  magnitude  of physical  processes,  which  are  driven by  galaxy
merging, should be proportional  to the merging galaxy number density.
Many phenomena have been linked  to galaxy merging, most notably, star
formation,  AGN activity,  and galaxy  morphologies.  The  cosmic star
formation rate density, as traced by radio, ultraviolet, far infrared,
and/or  various  emission  line  luminosities  has  been  measured  by
numerous             authors            at            $z\!\lesssim\!7$
\citep{lilly96,madau,con97,steid99,haa,brinch04,wyder05,bou06,hb06}.
While  there is  reasonable  consensus that  the  star formation  rate
density  increases  from $z\!\simeq\!0$  to  $z\!\simeq\!1$, the  high
redshift  portion   can  differ   greatly  among  the   many  authors.
Nonetheless, the peak in our  merger number density is consistent with
the   peak   in   the   cosmic   star  formation   rate   density   at
$z\!\simeq\!1\!-\!1.5$.   Moreover,  the   peak  star  formation  rate
density       is      $\sim\!10^{-1}$~M$_{\odot}$~yr$^{-1}$~Mpc$^{-3}$
\citep[eg.][]{hop06}, which would require the typical merger system to
have a star formation rate of $\sim\!100$~M$_{\odot}$~yr$^{-1}$.  This
rate is  considerably higher than  what is expected for  major mergers
\citep{bell05},  which suggests that  major mergers  are not  the only
source of  star formation at these redshifts.   However, minor merging
is expected to be more frequent, and may be responsible for triggering
the remaining portion of the cosmic star formation density.

Many  authors  have addressed  an  evolutionary  link between  merging
galaxies  and  AGN  \citep{c90,sr98,kh00,cro06,hop06},  and  generally
agree that  major merging  can drive AGN  activity.  Therefore,  it is
natural  to compare  our merger  number  density to  the observed  AGN
density.  The  density of type-1  AGN selected in  soft (0.5--2.0~keV)
$X$-ray          band           with          luminosities          of
$L_X\!=\!10^{42}-10^{46}$~erg~s$^{-1}$  from \citet{hasi} is  shown in
\fig{densfig}  as a  dotted line.   This sample  provides  the roughly
instantaneous AGN activity unobscured through any dust, and is similar
to the  results in the  hard $X$-rays \citep{ueda}.  This  AGN density
peaks  at   $z_{\rm  AGN}\!\simeq\!0.65$,   or  a  lookback   time  of
$\simeq\!5.9$~Gyr, whereas our observed merger number density peaks at
$z\!_{\rm dens}\!\simeq\!\zdens$,  or $\simeq\!\tmerg$~Gyr in lookback
time.  If  a major  merger of  two galaxies triggers  an AGN,  then we
expect on average $\sim\!\delt$~Gyr between the time when the galaxies
are  observed as a  pair on  the sky,  and when  the remnant  would be
observed as an AGN in the soft $X$-rays.  \citet{sdh05} have suggested
that major merging  can trigger the fueling of  the central black hole
after  the  merger  onset, and  will  be  observed  as a  visible  AGN
$\sim\!1\!-\!2$~Gyr later.  According  to these simulations, we expect
the merger  number density to  peak $\sim\!2.5$~Gyr earlier  in cosmic
time than the AGN density which is similar to our observations.

These  measurements have  exploited the  excellent  redshift accuracy,
spatial  resolution, and depth  of ACS  grism observations.   The Wide
Field  Camera~3  for  HST  will provide  combined  ultraviolet  prism,
infrared grism,  and numerous ultraviolet,  optical, and near-infrared
broadband  observations.  With  such  a  rich  dataset,  redshifts  of
comparable  accuracy from $2\!\lesssim\!z\!\lesssim\!6$  should become
standard.  This will be ideal  for refining this work, and determining
the role of various effects, such as galaxy environments and/or cosmic
variance at $z\!\lesssim\!6$.

\acknowledgments We  are very grateful for the  excellent comments and
suggestions of  the anonymous Referee which have  improved this paper.
We  additionally  thank  Rolf  Jansen, Norman  Grogin,  Nimish  Hathi,
Sangeeta  Malhotra, James  Rhoads, and  Nor  Pirzkal for  a number  of
helpful discussions.  RER acknowledges support from the ASU NASA Space
Grant and the ASU Graduate Student and Professional Association.  This
work was supported by grant HST-GO-15030.07-A from the Space Telescope
Science Institute, which  is operated by AURA under  NASA contract NAS
5-26555.


\begin{thebibliography}{}
\bibitem[Abraham et al.(1996)]{abr96} Abraham, R. G., Tanvir, N. R., Santiago, B. X., Ellis, R. G., Glazebrook, K., \& van den Bergh, S.  1996, \mnras, 279, L47
\bibitem[Abraham(1999)]{abr99} Abraham, R. G.  1999, IAUS, 186, 11
\bibitem[Bahcall et al.(1997)]{bah} Bahcall, J. N., Kirhakos, S., Saxe, D. H., \& Schneider, D. P.  1997, \apj, 479, 642
\bibitem[Beckwith et al.(2006)]{beck} Beckwith, S. V. W., et al. 2006, \aj, 132, 1729
\bibitem[Bell et al.(2005)]{bell05} Bell, E. F., et al. 2005, \apj, 625, 23
\bibitem[Bell et al.(2006)]{bell06} Bell, E. F., Phleps, S., Somerville, R. S., Wolf, C., Borch, A., \& Meisenheimer, K.  2006, \apj, 652, 270
\bibitem[Ben\'itez(2000)]{ben00} Ben\'itez, N. 2000, \apj, 536, 571
\bibitem[Berrier et al.(2006)]{ber06} Berrier, J. C., Bullock, J. S., Barton, E. J., Guenther, H. D., Zentner, A. R., \& Wechsler, R. H. 2006, \apj, 652, 56
\bibitem[Bershady, Jangren, \& Conselice(2000)]{ber00} Bershady, M. A., Jangren, A., \& Conselice, C. J.  2000, \aj, 119, 2645
\bibitem[Bolzonella, Miralles, \& Pell\'o(2000)]{hyperz} Bolzonella, M., Miralles, J.-M., \& Pell\'o, R. 2000, A\&A, 363, 476
\bibitem[Bouwens et al.(2006)]{bou06} Bouwens, R. J., Illingworth, G. D., Blakeslee, J. P., \& Franx, M.  2006, \apj, 653, 53
\bibitem[Brinchmann et al.(2004)]{brinch04} Brinchmann, J., Charlot, S., White, S. D. M., Tremonti, C., Kauffmann, G., Heckman, T., \& Brinkmann, J.  2004, \mnras, 351, 1151
\bibitem[Bruzual \& Charlot(2003)]{bc03} Bruzual, G. \& Charlot, S. 2003, \mnras, 344, 1000 (BC03)
\bibitem[Burkey et al.(1994)]{burk94} Burkey, J. M., Keel, W. C., Windhorst, R. A., \& Franklin, B. E.  1994, \apj, 429, 13
\bibitem[Canalizo et al.(2007)]{canal} Canalizo, G., Bennert, N., Jungwiert, B., Stockton, A., Schweizer, F., Lacy, M., \& Peng, C. 2007, \apj, 669, 801
\bibitem[Carlberg(1990)]{c90} Carlberg, R. G. 1990, \apj, 350, 505
\bibitem[Carlberg(1991)]{c91} Carlberg, R. G. 1991, \apj, 375, 429
\bibitem[Carlberg, Pritchet, \& Infante(1994)]{cpi94} Carlberg, R. G., Pritchet, C. J., \& Infante, L.  1994, \apj, 435, 540
\bibitem[Coe et al.(2006)]{coe} Coe, D., Ben\'itez, N., S\'anchez, S., F., Jee, M., Bouwens, R., \& Ford, H. 2006, \aj, 132, 926
\bibitem[Cohen et al.(2006)]{shc} Cohen, S. H., et al. 2006, \apj, 639, 731
\bibitem[Cole et al.(2001)]{cole} Cole, S., et al. 2001, \mnras, 326, 255
\bibitem[Coleman, Wu, \& Weedman(1980)]{cww} Coleman, G. D., Wu, C.-C., \& Weedman, D. W. 1980, \apjs, 43, 393
\bibitem[Connolly et al.(1997)]{con97} Connolly, A. J., Szalay, A. S., Dickinson, M., SubbaRao, M. U., \& Brunner, R. J.  1997, \apj, 486, L11
\bibitem[Conselice(2003)]{con03a} Conselice, C. J. 2003, \apjs, 147, 1
\bibitem[Conselice et al.(2003)]{con03} Conselice, C. J., Bershady, M. A., Dickinson, M., \& Papovich C.  2003, \apj, 126, 1183
\bibitem[Conselice(2006)]{con06} Conselice, C. J. 2006, \apj, 638, 686
\bibitem[Croton et al.(2006)]{cro06} Croton, D. J., et al. 2006, \mnras, 356, 1155
\bibitem[Daddi et al.(2004)]{dad04} Daddi, E., et al. 2004, \apj, 617, 746
\bibitem[Daddi et al.(2005)]{dad05} Daddi, E., et al. 2005, \apj, 626, 680
\bibitem[Daddi et al.(2007)]{dad07} Daddi, E., et al. 2007, \apj, in press, arXiv/0705.2832
\bibitem[De Lucia et al.(2006)]{del} De Lucia, G., Springel, V., White, S. D. M., Croton, D., \& Kauffmann, G. 2006, \mnras, 366, 499
\bibitem[Dickinson et al.(2003)]{d03} Dickinson, M., Papovich, C., Ferguson, H. C., \& Budav\'ari, T.  2003, \apj, 587,25
\bibitem[Driver et al.(1995)]{d95} Driver, S. P., Windhorst, R. A., Ostrander, E. J., Keel, W. C., Griffiths, R., E., Ratnatunga, K. U.  1995, \apj, 449, L23
\bibitem[Driver(1999)]{driv} Driver, S. P. 1999, \apj, 526, L69
\bibitem[Elsner, Feulner, \& Hopp(2007)]{elsn} Elsner, F., Feulner, G., \& Hopp, U.  2007, A\&A, accepted, astro-ph/0711.0384
\bibitem[Fakhouri \& Ma(2007)]{fak} Fakhouri, O. \& Ma, C.-P. 2007, \mnras, submitted, astro-ph/0710.4567
\bibitem[F\"orster Schreiber et al.(2004)]{fs04} F\"orster Schreiber, N. M., et al. 2004, \apj, 616, 40 
\bibitem[Grogin et al.(2005)]{grog} Grogin, N. A., et al. 2005, \apj, 627, L97
\bibitem[Guyon, Sanders, \& Stockton(2006)]{guy} Guyon, O., Sanders, D. B., \& Stockton, A. 2006, \apjs, 166, 89
\bibitem[Haarsma et al.(2000)]{haa} Haarsma, D. B., Partridge, R. B., Windhorst, R. A., \& Richards, E. A.  2000, \apj, 544, 641
\bibitem[Hasinger,  Miyaji,   \&  Schmidt(2005)]{hasi}  Hasinger,  G., Miyaji, G., \& Schmidt, M.  2005, A\&A, 441, 417
\bibitem[Hopkins \& Beacom(2006)]{hb06} Hopkins, A. M. \& Beacom, J. F. 2006, \apj, 651, 142
\bibitem[Hopkins et al.(2006)]{hop06} Hopkins, P. F., Hernquist, L., Cox, T. J., Di Matteo, T., Robertson, B., \& Springel, V.  2006, \apjs, 163, 1
\bibitem[Kartaltepe et al.(2007)]{kart07} Kartaltepe, J. S., et al. 2007, \apjs, 172, 320
\bibitem[Kauffmann \& Haehnelt(2000)]{kh00} Kauffmann, G. \& Haehnelt, M.  2000, \mnras, 311, 576
\bibitem[Kauffmann, White, \& Guiderdoni(1993)]{kwg93} Kauffmann, G., White, S. D. M., \& Guiderdoni, B.  1993, \mnras, 264, 201
\bibitem[Khochfar \& Burkert(2001)]{kb01} Khochfar, S. \& Burkert, A.  2001, \apj, 561, 517
\bibitem[Larson \& Tinsley(1978)]{lt78} Larson, R. B. \& Tinsley, B. M.  1978, \apj, 219, 46
\bibitem[Le~F\`evre et al.(2000)]{lefe00} Le F\`evre, O., et al.  2000, \mnras, 311, 565
\bibitem[Lilly et al.(1996)]{lilly96} Lilly, S. J., Le F\`evre, O., Hammer, F., \& Crampton, D.  1996, \apj, 406, L1
\bibitem[Lin et al.(2004)]{lin04} Lin, L., et al.  2004, \apj, 617, L9
\bibitem[Madau et al.(1996)]{madau} Madau, P., Ferguson, H. C., Dickinson, M. E., Giavalisco, M., Steidel, C. C., \& Fruchter, A.  1996, \mnras, 283, 1388
\bibitem[Neuschaefer et al.(1997)]{neu97} Neuschaefer, L. W., Im, W., Ratnatunga, K. U., Griffiths, R. E., \& Casertano, S.  1997, \apj, 480, 59
\bibitem[Oke \& Gunn(1983)]{abmag} Oke, J. B. \& Gunn, J. E. 1983, \apj, 266, 713
\bibitem[Papovich, Dickinson, \& Ferguson(2001)]{pap} Papovich, C., Dickinson, M., \& Ferguson, H. C. 2001, \apj, 559, 620
\bibitem[Patton et al.(1997)]{p97} Patton, D. R., Pritchet, C. J., Yee, H. K. C., Ellingson, E., \& Carlberg, R. G. 1997, \apj, 475, 29
\bibitem[Patton et al.(2000)]{p00} Patton, D. R., Carlberg, R. G., Marzke, R. O., Pritchet, C. J., da Costa, L. N., \& Pellegrini, P. S., 2000, \apj, 536, 153 (P2000)
\bibitem[Patton et al.(2002)]{p02} Patton, D. R., et al.  2002, \aj, 565, 208
\bibitem[Pirzkal et al.(2004)]{nor} Pirzkal,  N., et al.  2004, \apjs, 154, 501
\bibitem[de Propris et al.(2005)]{mgc} de Propris, R., Liske, J., Driver, S. P., Allen, P. D., \& Cross, N. J. G. 2005, \aj, 130, 1516
\bibitem[de Propris et al.(2007)]{dep07} de Propris, R., Conselice, C. J., Driver, S. P., Liske, J., Patton, D., Graham, A., \& Allen, P.  2007, astro-ph/0705.2528
\bibitem[Ryan et al.(2007)]{ryan07} Ryan Jr., R. E., et al.  2007, \apj, 668, 839
\bibitem[Salpeter(1955)]{salp} Salpeter, E. E. 1955, \apj, 121, 161
\bibitem[Sanders \& Mirabel(1996)]{sm96} Sanders, D. B. \& Mirabel, I. F.  1996, ARA\&A, 34, 749
\bibitem[Schechter(1976)]{sch} Schechter, P. 1976, \apj, 203, 297
\bibitem[Schneider et al.(2005)]{sdss} Schneider, D. N., et al.  2005, \aj, 130, 367
\bibitem[Silk \& Rees(1998)]{sr98} Silk, J. \& Rees, M. J. 1998, A\&A, 331, L1
\bibitem[Somerville et al.(2004)]{som} Somerville, R., S., Lee, K., Ferguson, H. C., Gardner, J. P., Moustakas, L., A., \& Giavalisco, M. 2004, 600, L171
\bibitem[Spergel et al.(2007)]{wmap} Spergel, D. N., et al.  2007, \apjs, 170, 377
\bibitem[Springel, Di Matteo, \& Hernquist(2005)]{sdh05} Springel, V., Di Matteo, T., \& Hernquist, L.  2005, \mnras, 361, 776
\bibitem[Steidel et al.(1996)]{st96} Steidel, C. C., Giavalisco, M., Pettini, M., Dickinson, M., \& Adelberger, K. L. 1996, \apj, 462, L17
\bibitem[Steidel et al.(1999)]{steid99} Steidel, C. C., Adelberger, K. L., Giavalisco, M., Dickinson, M., \& Pettini, M.  1999, \apj, 519, 1
\bibitem[Thompson et al.(2005)]{nic} Thompson, R. I., et al. 2005, \aj, 130, 1
\bibitem[Toomre(1977)]{t77} Toomre, A.  1977, in ``Evolution of Galaxies and Stellar Populations,'' ed. B. M. Tinsley \& R. B. Larson (New Haven: Yale Univ. Press), 401
\bibitem[Ueda et al.(2003)]{ueda} Ueda, Y., Akiyama, M., Ohta, K., \& Miyaji, T.  2003, \apj, 598, 886
\bibitem[White \& Frenk(1991)]{wf91} White, S. D. M. \& Frenk, C. S. 1991, \apj, 379, 52
\bibitem[Wyder et al.(2005)]{wyder05} Wyder, T. K., et al. 2005, \apj, 619, L15
\bibitem[Xu, Sun, \& He(2004)]{xu} Xu, C. K., Sun, Y. C., \& X. T. 2004, \apj, 603, L73
\bibitem[Yee \& Ellingson(1995)]{ye95} Yee, H. K. C. \& Ellingson, E.  1995, \apj, 445, 37
\bibitem[Zepf \& Koo(1989)]{zk89} Zepf, S. E. \& Koo, D. C. 1989, \apj, 337, 34
\end{thebibliography}
\end{document}